\documentstyle[11pt]{article}
\newlength{\bredde}
\def\slash#1{\settowidth{\bredde}{$#1$}\ifmmode\,\raisebox{.15ex}{/}
\hspace*{-\bredde} #1\else$\,\raisebox{.15ex}{/}\hspace*{-\bredde} #1$\fi}
\newcommand{\beq}{\begin{equation}}
\newcommand{\eeq}{\end{equation}}
\newcommand{\noi}{\vspace{12pt}\noindent}
\newcommand{\lG}{\raise.3ex\hbox{$\stackrel{\leftarrow}{G}$}}
\newcommand{\lU}{\raise.3ex\hbox{$\stackrel{\leftarrow}{U}$}}
\newcommand{\lP}{\raise.3ex\hbox{$\stackrel{\leftarrow}{{\cal P}}$}}
\newcommand{\leta}{\raise.3ex\hbox{$\stackrel{\leftarrow}{\eta}$}}
\newcommand{\lOmega}{\raise.3ex\hbox{$\stackrel{\leftarrow}{\Omega}$}}
\newcommand{\ldr}{\raise.3ex\hbox{$\stackrel{\leftarrow}{\delta^r}$}}

\def\beqn{\begin{eqnarray}}
\def\eeqn{\end{eqnarray}}

\def\gtwid{\raise.3ex\hbox{$>$\kern-.75em\lower1ex\hbox{$\sim$}}}
\def\ltwid{\raise.3ex\hbox{$<$\kern-.75em\lower1ex\hbox{$\sim$}}}

\def\la{\lambda}

\begin{document}
\topmargin -1.4cm
\oddsidemargin -0.8cm
\evensidemargin -0.8cm
\title{\Large{{\bf Consistency Conditions for Finite-Volume
Partition Functions}}}

\vspace{1.5cm}

\author{~\\{\sc G. Akemann}\\Centre de Physique Th\'eorique CNRS\\
Case 907 Campus de Luminy\\F-13288 Marseille Cedex 9\\France\\~\\
{\sc P.H. Damgaard}\\
The Niels Bohr Institute\\ Blegdamsvej 17\\ DK-2100 Copenhagen {\O}\\
Denmark}
\date{} 
\maketitle
\vfill
\begin{abstract}Using relations from random matrix theory, we derive
exact expressions for all $n$-point spectral correlation functions
of Dirac operator eigenvalues in terms of finite-volume partition functions. 
This is done for both chiral symplectic and chiral unitary
random matrix ensembles, which correspond to $SU(N_c \geq 3)$ gauge 
theories with $N_f$ fermions in the adjoint and fundamental
representations, respectively. In the latter case we infer from this an 
infinite sequence of consistency conditions that must be satisfied by the
corresponding finite-volume partition functions. 
\end{abstract}
\vfill

\begin{flushleft}
CPT-98/P.3617\\
NBI-HE-98-05 \\
hep-th/9802174
\end{flushleft}
\thispagestyle{empty}
\newpage

The computation of finite-volume partition functions, and in particular the
finite-size scaling of Dirac-operator eigenvalue correlations, has
been very elegantly phrased in terms of certain random matrix theory
distributions \cite{V} which have turned out to be universal 
\cite{ADMN,DN,SV}. This has led to a highly increased understanding
of field theories with spontaneous chiral symmetry breaking (such as QCD)
in what is called the mesoscopic scaling regime of finite volumes, a study
that in this particular context was initiated by the work of Leutwyler
and Smilga \cite{LS} (see also ref. \cite{SmV} for generalizations). The
central idea is that a scaling region exists in which the 
correlations of rescaled Dirac eigenvalues are exactly computable. As the
volume is taken to infinity, the scale of magnification is correspondingly
increased. There is already evidence from lattice gauge theory simulations
\cite{BBMSVW} that universal, exact, scaling functions are reached in
this limit.

\noi
Recently, there has been a flurry of activity related to both proofs
of universality \cite{ADMN,SV}, and in general to the extension of these
results to the double-microscopic scaling regime in which both fermion
masses and Dirac eigenvalues are rescaled at the same rate
in the large-volume limit \cite{DN,D,WGW}. One essential observation
in this connection is that the relevant Dirac eigenvalue distributions
are also computable directly from the field theoretic 
finite-volume partition functions, without having to go through the 
random matrix theory formulation \cite{D1,AD}. The new ingredient needed
is a knowledge of finite-volume partition functions with additional
fermion species, the masses of these additional fermions taking the
r\^{o}les of Dirac eigenvalues in the original theory.

\noi
In this paper we shall derive a novel set of relations which provide
double-microscopic spectral correlators in terms of suitably extended 
finite-volume partition functions. Our tool shall again be random
matrix theory, but, as before \cite{D1,AD}, the final expressions will involve
only finite-volume partition functions, without reference to 
random matrix theory. We shall do this for both the cases corresponding
to chiral symplectic random matrix theory ensembles ($SU(N_c\!\geq\!3)$
gauge theories coupled to $N_f$ fermions in the adjoint representation),
and those corresponding to chiral unitary ensembles ($SU(N_c\!\geq\!3)$
gauge theories coupled to $N_f$ fermions in the fundamental representation).
Surprisingly, in the latter case the expressions we get are very different
from those one obtains by using factorization of correlation functions
in terms of the unitary kernel (which also can be represented directly
in terms of extended partition functions). 
In fact, while the factorization formula
shows that all higher correlation functions can be obtained by means of
the kernel (which has been shown to be related to the partition function
with just two additional fermion species \cite{D1}), the new relations
involve in this case the partition functions with, for $k$-point correlation
functions, $2k$ additional fermions. This in turn implies highly stringent
consistency conditions these finite-volume partition functions must
satisfy.

\noi
Before we turn to the consistency conditions, we first describe the
derivation of new relations between Dirac eigenvalue correlators and
finite-volume partition functions. Our starting point is the (chiral) random
matrix formulation \cite{V}:
\beq
\tilde{\cal Z}_{\nu}^{(N_{f},\beta)}(m_1,\ldots,m_{N_{f}}) 
~=~ \int\! dW \prod_{f=1}^{N_{f}}{\det}\left(iM + m_f\right)~
\exp\left[-\frac{N\beta}{4} \mbox{tr}\, V(M^2)\right] ~,\label{zMch}
\eeq
with
\beq
M ~=~ \left( \begin{array}{cc}
              0 & W^{\dagger} \\
              W & 0
              \end{array}
      \right) ~,
\eeq    
and where $\beta$ from now on labels the matrix ensemble. Thus, $\beta\!=\!4$
corresponds to the symplectic ensemble, and $\beta\!=\!2$ to the unitary
ensemble. The matrices $W$ are rectangular complex matrices of size
$N\times(N\! +\! \nu)$, and they are integrated over with the Haar measure.
The space-time volume $V$ of the finite-volume gauge theory is, in the 
large-$N$ limit, identified with $2N$. The topological index $\nu$, for
convenience always taken to be non-negative here, also counts the number of 
zero modes of the matrix $M$.

\noi
Written in terms of the eigenvalues $\la_i$ of the hermitian matrix 
$W^{\dagger}W$, the partition functions 
$\tilde{\cal Z}_{\nu}^{(N_{f},\beta)}$ are (ignoring 
unimportant overall factors): 
\beq
\tilde{\cal Z}_{\nu}^{(N_{f},\beta)}(m_1,\ldots,m_{N_{f}})  ~=~ 
\prod_{f=1}^{N_{f}} 
(m_f^{\nu})\int_0^{\infty}\! \prod_{i=1}^N \left(d\lambda_i 
~\la_i^{\frac{\beta}{2}\nu + \frac{\beta}{2}-1}~
\prod_{f=1}^{N_{f}}(\lambda_i + m_f^2)~
\mbox{e}^{-\frac{N\beta}{2}V(\lambda_i)}\right)\left|{\det}_{ij}
\lambda_j^{i-1}\right|^{\beta} ~.\label{zmatrixeigen}
\eeq

\noi
We will treat the chiral unitary ($\beta\!=\!2$) and the chiral symplectic
($\beta\!=\!4$) ensembles in the same fashion in what follows. In principle
most of the relations carry over to the chiral orthogonal ($\beta\!=\!1$)
ensemble (which corresponds to $SU(2)$ gauge theory with $N_f$ fermions
in the fundamental representation) as well, but the final steps where we 
identify finite-volume
field theory partition functions require $\beta$ to be even. For this
reason $\beta$ is restricted to the values 2 and 4 in the following.

\noi
Let us define
\beq
\rho^{(N_f,\nu,\beta)}(\la_1,\ldots,\la_N;m_1,\ldots,m_{N_f})
~\equiv~ \frac{1}{\tilde{\cal Z}_{\nu}^{(N_{f})}(\{m_f\})}
\prod_f^{N_f}(m_f^\nu)
\prod_i^N w_{\beta}(\la_i) \prod_{j<l}^N|\la_j-\la_l|^{\beta} ~,
\label{densgen}
\eeq
where
\beq
w_{\beta}(\la) ~=~ \la^{\frac{\beta}{2}\nu+\frac{\beta}{2}-1}
\prod_f^{N_f}(\la+m_f^2)\mbox{e}^{-\frac{N\beta}{2}V(\la)} ~.
\label{w}
\eeq
Definition (\ref{densgen}) is proportional to the integrand of the
partition function eq. (\ref{zmatrixeigen}). In the last term we have
rewritten the Vandermonde determinant in the standard way.
All correlation functions with $k\!<\!N$ can now be obtained from the 
density of eq. 
(\ref{densgen}) by integrating out a suitable number of eigenvalues:
\beqn
\rho^{(N_f,\nu,\beta)}(\la_1,\ldots,\la_k;\{m_f\}) &=&
\int_0^\infty\prod_{i=k+1}^N( d\la_i)\
\rho^{(N_f,\nu,\beta)}(\la_1,\ldots,\la_N;m_1,\ldots,m_{N_f}) \nonumber\\
&=& \frac{1}{\tilde{\cal Z}_{\nu}^{(N_{f})}(\{m_f\})}\prod_f^{N_f}(m_f^\nu)
\prod_i^k w_{\beta}(\la_i) \prod_{j<l}^k|\la_j-\la_l|^{\beta} \nonumber\\
&&\times\int_0^{\infty}\! \prod_{i=k+1}^N
\left(\ d\la_i\ w_{\beta}(\la_i)
\prod_{j=1}^k|\la_i+(i\sqrt{\la_j})^2|^{\beta}\right)
\prod_{k+1\leq j<l\leq N} |\la_j-\la_l|^{\beta}\nonumber\\
&=& \prod_i^k\left( (i\sqrt{\la_i})^{-\beta\nu}w_{\beta}(\la_i)\right)
\prod_{j<l}^k|\la_j-\la_l|^{\beta}
\frac{\tilde{\cal Z}_{\nu}^{(N_{f}+\beta k)}(\{m_f\};\{i\sqrt{\la_j}\})}
{\tilde{\cal Z}_{\nu}^{(N_{f})}(\{m_f\})} ~.
\label{corrmm}
\eeqn
In the first step of the calculation
we have taken the weight functions $w_{\beta}(\la_{i=1,\ldots,k})$
out of the integral and split the Vandermonde determinant
into a prefactor, the additional
mass terms\footnote{This is the precise point where the considerations
do not immediately carry over the $\beta\!=\!1$ case, since in that case
we cannot disregard the absolutely value of the Vandermonde determinant.
There is therefore no immediately obvious way of writing it directly
in terms of massive partition functions.}, and a remaining Vandermonde 
determinant.
In the second step we have disregarded, in the large-$N$ limit,
the difference between the integral for $N-k$ eigenvalues and the matrix 
model partition function of $N$ eigenvalues. It contains $\beta k$ additional
imaginary masses
$i\sqrt{\la_1},i\sqrt{\la_2},
\ldots,i\sqrt{\la_k}$, each of which is $\beta$-fold degenerate.

\noi
We now go back to the original picture, in which we seek 
correlators of eigenvalues $z_i$ of the Dirac operator, with 
$\la_i\!=\!z_i^2$. 
We also go to the double-microscopic limit in which $\zeta_i \equiv
z_i N2\pi\rho(0)$ and $\mu_f \equiv m_f N2\pi\rho(0)$ are kept fixed as
$N\!\to\!\infty$. All factors of 
$\exp[-\frac{N\beta}{4}V(\zeta^2)]$ 
in the measure $w_{\beta}(\zeta_i^2)$ standing
outside the integral in (\ref{corrmm}) become replaced by unity in this 
limit, and by
identifying $\Sigma = 2\pi\rho(0)$, we can now compare
with the field theory finite-volume partition functions.
We then obtain the following expression for the
density correlators of the scaled $\zeta_i$-variables:
\beqn
\rho_S^{(N_f,\nu,\beta)}(\zeta_1,\ldots,\zeta_k;\mu_1,\ldots,\mu_{N_f}) 
&=& C_{\beta}^{(k)} 
\prod_i^k\left( \zeta_i^{\beta -1}
\prod_f^{N_f}(\zeta_i^2+\mu_f^2)\right)
\prod_{j<l}^k|\zeta_j^2-\zeta_l^2|^{\beta} \nonumber\\
&&\times\
\frac{{\cal Z}_{\nu}^{(N_{f}+\beta k)}
(\mu_1,\ldots,\mu_{N_f};\{i\zeta_1\},\ldots, \{i\zeta_k\})}
{{\cal Z}_{\nu}^{(N_{f})}(\mu_1,\ldots,\mu_{N_f})} ,
\label{corrft}
\eeqn
where each additional mass $i\zeta_j$ is $\beta$-fold degenerate.
The overall proportionality constant $C_{\beta}^{(k)}$
is of course not given {\em a priori},
and has to be fixed by a matching condition. Moreover, the proportionality
constant could in principle depend on $k$, as indicated.

\noi
Let us now consider the cases $\beta\!=\!4$ and $\beta\!=\!2$ separately,
beginning with the case of $\beta\!=\!4$. As we have explained
elsewhere \cite{AD}, a
subset of the spectral correlations derived above also follow from some
of the general theorems that have been proven by Mahoux and Mehta \cite{MM}
using the quaternion formalism. The difficulty there is that the
quantity $f_4(\la_i,\la_j)$, which corresponds to the kernel of the 
skew-orthogonal polynomials, is now a {\em quaternion}. It can be represented 
by a $2\!\times\!2$ matrix. The correlation functions of eigenvalues are then
given by quaternion determinants $\det[f_4(\la_i,\la_j)]_m$ of the kernel
$f_4(\la_i,\la_j)$. We have not been able to express this kernel itself
in terms of matrix model (and thus, in the double-microscopic scaling limit, 
finite-volume field theory)
partition functions, but one can easily express the determinant of this
kernel in terms of partition functions. For instance, using Theorem 1.2
of ref. \cite{MM} we are immediately led to eq. (\ref{corrft}) for $k\!=\!1$,
and also the density-density correlator ($k\!=\!2$) can be derived in 
an analogous way \cite{AD}. But the relation (\ref{corrft}) is of course
far more general.

\noi
The chiral unitary ensemble (eq. (\ref{corrft}) with $\beta\!=\!2$) 
is actually
at present far more interesting, since we in that case already have an
alternative description of the same spectral correlators.
This is summarized by the master formula for the kernel \cite{D1,AD},
\beq
K_S^{(N_{f},\nu)}(\zeta,\zeta';\mu_1,\ldots,\mu_{N_{f}}) ~=~ 
(-1)^{\nu+[N_f/2]}\sqrt{\zeta\zeta'}\prod_{f}^{N_{f}}
\sqrt{(\zeta^2+\mu_f^2)(\zeta'^2+\mu_f^2)}~\frac{
{\cal Z}_{\nu}^{(N_{f}+2)}(\mu_1,\ldots,\mu_{N_{f}},i\zeta,i\zeta')}{
{\cal Z}_{\nu}^{(N_{f})}(\mu_1,\ldots,\mu_{N_{f}})} \label{mf}
\eeq
from which all higher $k$-point correlation functions follow:
\beq
\rho_S^{(N_{f},\nu)}(\zeta_1,\ldots,\zeta_k;\mu_1,\ldots,\mu_{N_{f}}) ~=~
\det_{1\leq a,b\leq k} 
K_S^{(N_{f},\nu)}(\zeta_a,\zeta_b;\mu_1,\ldots,\mu_{N_{f}}) ~.
\label{correlchUE}
\eeq
A quick glance reveals that these two description (eq. (\ref{corrft}) for
$\beta\!=\!2$, and eq. (\ref{correlchUE})) are very different for $k\!\neq\!1$.
For $k\!=\!1$ the two expressions agree up to the overall constant 
$C^{(1)}_2$, which thus is fixed in that case:
\beq
C^{(1)}_2 ~=~ (-1)^{\nu+[N_f/2]} ~.
\eeq

\noi
For higher $k$-point correlation functions, the two alternative descriptions
imply non-trivial consistency conditions for the partition functions
involved. Surprisingly, we see that these conditions must relate the 
finite-volume partition functions {\em with a different number of fermion 
species} to each other. The relations become particularly transparent
if we first analytically continue the additional (``fictitious'') fermions
masses onto physical values by $\zeta_j\to -i\zeta_j$. Then we immediately
obtain the following infinite sequence of consistency conditions:
\begin{eqnarray}
&&\det_{1\leq a,b\leq k}\left[\sqrt{\zeta_a\zeta_b}\prod_{f=1}^{N_{f}}
\sqrt{(\mu_f^2-\zeta_a^2)(\mu_f^2-\zeta_b^2)}~
{\cal Z}_{\nu}^{(N_{f}+2)}(\mu_1,\ldots,\mu_{N_{f}},\zeta_a,\zeta_b)\right]
= \cr && C_2^{(k)}(-1)^{k(\nu+[N_f/2]+1)}
\prod_i^k\left( \zeta_i\prod_{f=1}^{N_{f}}
(\mu_f^2-\zeta_i^2)\right)
\prod_{j<l}^k|\zeta_j^2-\zeta_l^2|^2 ~
\frac{{\cal Z}_{\nu}^{(N_{f}+2k)}
(\mu_1,\ldots,\mu_{N_f},\{\zeta_1\},\ldots,\{\zeta_k\})}
{{\cal Z}_{\nu}^{(N_{f})}(\mu_1,\ldots,\mu_{N_{f}})^{1-k}} \nonumber\\
\label{cons}
\end{eqnarray}
In this case the finite-volume partition functions are thus highly constrained
by relations that link theories with $N_f\!+\!2k$ fermions to those of
$N_f\!+\!2$ and $N_f$ fermions.
These relations become quite involved for increasing values
of $k$. There are known exact expressions for the finite-volume partition
functions for this case, which corresponds to $SU(N_c\!\geq\!3)$ gauge 
theories with $N_f$ fermions in the fundamental representation \cite{JSV}:
\beq
Z_{\nu}^{(N_{f})}(\mu_1,\ldots,\mu_{N_{f}}) ~=~ 
\frac{\det {\cal A}(\{\mu_i\})}{\Delta(\mu^2)} ~, \label{Z}
\eeq
where the $N_f\!\times\! N_f$ matrix ${\cal A}(\{\mu_i\})$ conveniently
can be written \cite{D1}
\beq
{\cal A}_{ij} ~=~ \mu_i^{j-1}I_{\nu+j-1}(\mu_i) ~,\label{Aconv}
\eeq
and where $\Delta(\mu^2)$ stands for the Vandermonde determinant of the 
squared masses $\mu_i^2$. Using this explicit representation, we have 
verified in a few of the
simpler cases that these consistency conditions indeed are satisfied.

\noi
One of the surprising consequences of the connection to random matrix theory 
is that the finite-volume field theory partition functions can be used
directly to compute the universal double-microscopic limits of those
orthogonal polynomials that are associated with the random matrix technique
\cite{AD}. While there at the moment is no obvious interpretation of these
orthogonal polynomials in field theory terms, it is interesting to note
that the connection between the kernel 
\beqn
K_S^{(N_{f},\nu)}(\zeta,\zeta';\{\mu_f\}) &=&
C_2(\zeta\zeta')^{\nu+\frac{1}{2}}\prod_f^{N_f}
\sqrt{(\zeta^2+\mu_f^2)(\zeta'^2+\mu_f^2)}\frac{1}{\zeta^2-\zeta'^2}\nonumber\\
&\times& \left[ P_{N-1}^{(N_{f},\nu)}(\zeta^2;\{\mu_f\})
P_{N}^{(N_{f},\nu)}(\zeta'^2;\{\mu_f\}) -
P_{N}^{(N_{f},\nu)}(\zeta^2;\{\mu_f\})
P_{N-1}^{(N_{f},\nu)}(\zeta'^2;\{\mu_f\})\right] \nonumber\\
\label{KOP}
\eeqn
and these orthogonal polynomials provide us with yet more consistency
conditions that must be imposed on the finite-volume partition functions.
We have already reproduced the relation (\ref{mf}) between the kernel and the
partition functions. We now compare this with the corresponding relation
for the double-microscopic limit of the orthogonal polynomials \cite{AD},
\beq
P_{N}^{(N_{f},\nu)}(\zeta^2;\mu_1,\ldots,\mu_{N_{f}}) ~=~ C_3 
(-1)^N(i\zeta)^{-\nu}\
\frac{{\cal Z}^{(N_{f}+1)}_\nu(\mu_1,\ldots,\mu_{N_{f}},i\zeta)}
{{\cal Z}^{(N_{f})}_\nu(\mu_1,\ldots,\mu_{N_{f}})} ~, \label{polzch}
\eeq 
where the normalization constant $C_3$ is left unspecified.\footnote{This
overall constant simply specifies the normalization of the polynomials,
and we easily fix it once we choose the prescription (monic, or otherwise).
However, there is no need to fix this constant here.} Inserting
eq. (\ref{polzch}) into eq. (\ref{KOP}) and expanding in $1/N$ we obtain
the following set of consistency conditions
\beqn
{\cal Z}_\nu^{(N_f+2)} (\{\mu_f\},\zeta,\zeta') &=& 
\frac{C}{(\zeta'^2-\zeta^2){\cal Z}_\nu^{(N_f)} (\{\mu_f\})} \nonumber\\
&&\times \left[ \left( (\sum_f^{N_f}\mu_f\partial_{\mu_f} 
+\zeta\partial_{\zeta})
{\cal Z}_\nu^{(N_f+1)}(\{\mu_f\},\zeta)\right)
{\cal Z}_\nu^{(N_f+1)}(\{\mu_f\},\zeta') ~-~ 
\left(\zeta \leftrightarrow \zeta'\right)\right] \nonumber\\
 \label{cons2}
\eeqn
where we have again rotated back to real fermion masses.
There is yet another relation from random matrix models among the 
orthogonal polynomials themselves, which relates the polynomials
with $N_f$ massive flavors to those with $N_f+1$ \cite{DN}.
Surprisingly enough this relation leads to precisely the same consistency
conditions eq. (\ref{cons2}). One can fix the proportionality constant
by tracing it back to the matching condition between the double-microscopic
spectral density (the kernel evaluated at coincident points), but we leave
it here unspecified since the proportionality of the left and right hand sides
of eq. (\ref{cons2}) already gives a highly non-trivial series of conditions.
Using the explicit expression eq. (\ref{Z}) we have verified
in the first few cases that the relations of  eq. (\ref{cons2}) are satisfied.
Taking the consistency conditions eqs. (\ref{cons}) and (\ref{cons2})  
together the finite-volume partition functions for theories
with $N_f\!+\!2k$ fermions are now given only in terms of those of
$N_f\!+\!1$ and $N_f$ fermions.

\noi
The results presented above trivially carry over from the chiral unitary
ensemble to the ordinary unitary ensemble, which has been conjectured to
describe $SU(N_c\!\geq\!3)$ gauge theories with an {\em even} number
of fermions $N_f$ in $(2\!+\!1)$ dimensions \cite{V}.
The partition function of that ensemble is
\beq
\tilde{\cal Z}^{(N_{f})}(m_1,\ldots,m_{N_{f}})
~=~ \int\! dM \prod_{f=1}^{N_{f}}{\det}\left(M + im_f\right)~ 
\exp(-N\mbox{tr}\, V(M^2)) ~,
\label{ZUE}
\eeq
where the integration is over the Haar measure of 
hermitian $N\times N$ matrices $M$, and where masses are grouped into
pairs of opposite signs:
$$ {\mbox{\rm diag}}
(m_1,m_2,\ldots,m_{N_{f}/2},-m_1,-m_2,\ldots,-m_{N_{f}/2}) ~.
$$
In terms of the eigenvalues $\lambda_i$ of the hermitian
matrix $M$ this gives:
\beq
\tilde{\cal Z}^{(N_f)}(m_1,\ldots,m_{N_{f}}) ~=~
\int_{-\infty}^{\infty}\! \prod_{i=1}^N \left(d\lambda_i
\prod_{f=1}^{N_f/2}(\lambda_i^2 + m_f^2)~
\mbox{e}^{-NV(\lambda_i^2)}\right)\left|{\det}_{ij}
\lambda_j^{i-1}\right|^2 ~,
\eeq
where we again have ignored all irrelevant overall factors.

\noi
We then immediately get an analogous sequence of consistency conditions
for the finite-volume partition functions of this ensemble. The kernel
in this case follows from the master formula \cite{AD}
\beq
K_S^{(N_{f})}(\zeta,\zeta';\mu_1,\ldots,\mu_{N_{f}}) ~=~
\frac{1}{2\pi}
\prod_{f}^{N_{f}/2}\sqrt{(\zeta^2+\mu_f^2)(\zeta'^2+\mu_f^2)}~\frac{
{\cal Z}^{(N_{f}+2)}(\mu_1,\ldots,\mu_{N_{f}},i\zeta,i\zeta')}{
{\cal Z}^{(N_{f})}(\mu_1,\ldots,\mu_{N_{f}})} ~.\label{mfUE}
\eeq
The analogue of eq. (\ref{cons}) therefore becomes 
\begin{eqnarray}
&&\det_{1\leq a,b\leq k}\left[\prod_{f=1}^{N_{f}/2}
\sqrt{(\mu_f^2-\zeta_a^2)(\mu_f^2-\zeta_b^2)}~
{\cal Z}^{(N_{f}+2)}(\mu_1,\ldots,\mu_{N_{f}},\zeta_a,\zeta_b)\right]
= \cr && \tilde{C}^{(k)}(2\pi)^k\prod_i^k\left(\prod_{f=1}^{N_{f}/2}
\left(\mu_f^2-\zeta_i^2\right)\right)
\prod_{j<l}^k|\zeta_j-\zeta_l|^2 
\frac{{\cal Z}^{(N_{f}+2k)}
(\mu_1,\ldots,\mu_{N_f},\{\zeta_1\},\ldots,\{\zeta_k\})}
{{\cal Z}^{(N_{f})}(\mu_1,\ldots,\mu_{N_{f}})^{1-k}} ~.
\label{consUE}
\end{eqnarray}
The representation of the involved finite-volume field theory partition
functions in terms of group manifold integrals was given by Verbaarschot
and Zahed in the third paper of ref. \cite{V}, and explicitly worked 
out in ref. \cite{DN} (for $N_f$ even). A relation similar to eq.
(\ref{cons2}) between partition functions with odd and even $N_f$
can be worked out as well. The corresponding orthogonal
polynomials in terms of partition functions have been given in ref.
\cite{AD}.

\noi
To conclude: We have extended the analysis of refs. \cite{D1,AD} to the
case of higher $k$-point correlation functions of Dirac eigenvalues in
terms of finite-volume partition functions for $SU(N_c\!\geq\!3)$
gauge theories coupled to $N_f$ fermions in both the fundamental and adjoint
representations. For the case of adjoint fermions, relations for 
higher $k$-point spectral correlators 
in terms of finite-volume partition functions are new.
For the case of fundamental fermions, we have made the derivation 
without going through the factorization formalism based on the (chiral)
unitary kernels. This has allowed us to establish infinite sequences
of consistency conditions for the involved partition functions. 
We have also shown how a different sequence of consistency conditions arise
from comparing the expression for the orthogonal polynomials with that of
the kernel. This new set of consistency conditions involves both the partition
functions themselves, and their derivatives.
All of these relations share the remarkable property of being easily derived
on the basis of the connection to random matrix theory, while their origin
in proper field theory terms remain obscure at present. It is a
challenge to explain these relations at the level of effective
Lagrangians in the finite-volume ``mesoscopic'' scaling regime.

\vspace{1.5cm}
\noi
{\sc Acknowledgment:}\\
The work of G.A. is supported by European Community
grant no. ERBFMBICT960997, and the work of P.H.D. is partially supported by
EU TMR grant no. ERBFMRXCT97-0122.

\end{document}